\documentclass[aps,superscriptaddress,twocolumn,preprintnumbers,showkeys,showpacs,amsmath,amssymb]{revtex4}
\usepackage[cp1251]{inputenc}
\usepackage[english]{babel}
\usepackage{amssymb,amsmath}
\usepackage{amstext,textcomp} 
\usepackage[dvips]{hyperref}
\usepackage[mathscr]{eucal}
\usepackage{longtable}
\usepackage{graphicx}
\begin{document}
\title{Symmetry of Differential Equations and Quantum Theory}
\author{Dmitri Yerchuck (a),  Alla  Dovlatova (b), Andrey Alexandrov (b)\\
\textit{(a) Heat-Mass Transfer Institute of National Academy of Sciences of RB, Brovka Str.15, Minsk, 220072, dpy@tut.by\\ 
(b) M.V.Lomonosov Moscow State University, Moscow, 119899}}
\date{\today}%
\begin{abstract} The symmetry study of main differential equations of mechanics and electrodynamics  has shown, that differential equations, which
are invariant under transformations of groups, which are symmetry groups of mathematical
numbers (considered within the frames of the number theory) determine the mathematical nature of the
quantities, incoming in given equations. It allowed to proof the main postulate of quantum mechanics, consisting in that, that to any mechanical
quantity can be set up into the correspondence the Hermitian matrix by quantization. 

High symmetry of Maxwell equations allows to show, that to quantities, incoming in given equations can be set up into the correspondence the Quaternion (twice-Hermitian) matrix by their quantization.

It is concluded, that the equations of the dynamics of mechanical systems are not invariant under transformations
of quaternion multiplicative group and, consecuently, direct application of quaternions with usually used basis \{e, i, j, k \}  to build the new version of quantum mechanics, which was undertaken in the number of modern publications, is incorrect. It is the consequence
of non-abelian character of given group.  At the same time  we have found the correct ways for the creation of the new versions of quantum mechanics on the quaternion base by means of choice of new bases in quaternion ring, from which can be formed the bases for complex numbers under multiplicative groups  of which the equations of the dynamics of mechanical systems are  invariant.

\end{abstract}  

\maketitle \section{Introduction}
It is well known, that quantum theory is formulated by using of some postulates. For instance, the study of quantum mechanics in all textbooks (see, for example, \cite{Landau_L.D}, \cite{Davydov}) is beginning from the formulation of the main postulate, which is the following: "The linear self-adjoint operators are setting up into correspondence to observable physical quantities". It is the first main postulate of quantum mechanics. Given postulate is in fact containing in the first works on  quantum mechanics, which are referred to be the birth of quantum mechanics, that is in the papers of Heisenberg, Born, Jordan, Dirac, Pauli.
 The mathematical basis of Heisenberg's treatment \cite{Heisenberg} is the law of multiplication of quantum-theoretical quantities, which he derived 
from an ingenious consideration of correspondence arguments. The 
development of his formalism, which has been given in the paper of Born and Jordan \cite{Born}, is based upon the 
fact that this rule of multiplication is none other than the well-known 
mathematical rule of matrix multiplication. The infinite square array 
(with discrete or continuous indices) termed a matrix, is a representation of a physical quantity 
which is given in classical theory being to be  a function of time. The  
mathematical method of treatment inherent in the quantum mechanics 
is thereby characterized through the employment of matrix analysis 
in place of the usual number analysis \cite{Born}. It is clear, that matrices realise the representation of corresponding operators in corresponding space.
Moreover, it is directry formulated, that, the matrices, which are setting up in the correspondence  to observable physical quantities have to be Hermitian  matrices. It is given in \cite{Born} on concrete example of  the description of the dynamic system.
The arguments for introduction of the first main postulate were determined by the possibility to explain the experimental data, which could not be explained by classical theory. It is  in particular the interpretation of the spectrum of hydrogen atom which has been done for the first time by Pauli \cite{Pauli_M}.
The significant moment for correctness of the introduction of given postulate was the the possibility of connection of quantum mechanics with classical mechanics. Given requirement  is especially characteristic for the work of Dirac \cite{Dirac_P.A.M}. Dirac re-discovered and developed in given paper the main ideas of the Born-Jordan paper. For instance, starting with classical mechanics, Dirac 
writes the quantum  equation of motion for any quantity $x$
 \begin{equation}
\label{eq4A}  
\frac{dx}{dt} = [x,\mathcal{H}], 
\end{equation} 
where $\mathcal{H}$ is Hamiltonian. The quantum mechanics developed in \cite{Born}  from 
Heisenberg's approach was extended in \cite{Born_Heisenberg} to systems having arbitrarily many-degrees of freedom using the same main postulate.

  In Heisenberg's fundamental paper  and in the papers \cite{Born}, \cite{Born_Heisenberg},\cite{Pauli_M},\cite{Dirac_P.A.M}
following it, dynamical variables were assumed to be represented by 
matrices. In paper \cite{DiracP.A.M}, Dirac starts with a more general  
assumption. He supposes the quantum variables $x, y, ...$ to be elements of an 
algebra, which means that sums $x+y$ and products $x y$ are defined 
which satisfy the ordinary laws of algebra, excluding the commutative 
law of multiplication. Among the elements of this algebra are the 
numbers of classical mathematics which were called $c$-numbers. All 
other elements were called $q$-numbers. It was in fact  the step to the proof of the first main postulate, although the attempt to proof given postulate was not undertaken from the birth of quantum mechanics in 1925 up to now. We will show in present report, that given postulate can be  strictly proved.

\section{Algebra of the complex and hypercomplex  numbers and its connection with the symmetry of quantum systems}

Some useful results from algebra of the complex numbers were  summarized in the paper \cite{A_Dovlatova_D_Yerchuck}. They will be reproduced here for convenience of the readers and will be supplemented with some  results from algebra of the  hypercomplex  numbers. 

The numbers $1$ and $i$ are usually used to be basis of the linear space of complex numbers over the field of real numbers. At the same time to any complex number $a + ib$ can be set up in conformity the $[2 \times 2]$-matrix according to biective mapping
\begin{equation}
\label{eq1abcd}
 f : a + ib \to \left[\begin{array} {*{20}c} a&-b  \\ b&a \end{array}\right].
\end{equation}
Bijectivity of mapping (\ref{eq1abcd}) indicates on the existence of inverse mapping, that is to any matrix, which has the structure, given by right side in relation (\ref{eq1abcd}), corresponds the complex number, determined by left side. 
 The matrices
\begin{equation}
\label{eq61}
\left[\begin{array} {*{20}c} 1&0  \\ 0&1 \end{array}\right], 
\left[\begin{array} {*{20}c} 0&-i  \\ i&0 \end{array}\right]
\end{equation}
produce basis for complex numbers $\{a + ib\}$, $a,b \in R$  in the linear space  of $[2 \times 2]$-matrices, defined over the field of real numbers.

 It is convenient often to define the space of complex numbers over the group of real positive numbers, then the dimensionality of the matrices and basis has to be duplicated, since to two unities - positive $1$ and negative $-1$ 
can be set up in conformity the $[2 \times 2]$-matrices according to biective mapping
\begin{equation}
\label{eq62}
 \xi : 1 \to \left[\begin{array} {*{20}c} 1&0  \\ 0&1  \end{array}\right], -1 \to \left[\begin{array} {*{20}c} 0&1  \\ 1&0 \end{array}\right],
\end{equation}
 which allow to recreate the  operations with negative numbers without recourse of negative numbers themselves. Consequently, the following $[4 \times 4]$-matrices, so called [0,1]-matrices, can be basis of complex numbers
\begin{equation}
\label{eq63}
\begin{split}
\raisetag{40pt}
&\zeta : 1 \to [e_1]=\left[\begin{array} {*{20}c}1&0&0&0 \\  0&1&0&0  \\ 0&0&1&0 \\  0&0&0&1 \end{array}\right], \\
&i \to  [e_2]=\left[\begin{array} {*{20}c}  0&1&0&0  \\ 0&0&1&0 \\0&0&0&1\\ 1&0&0&0 \end{array}\right],\\
&-1 \to [e_3]=\left[\begin{array} {*{20}c} 0&0&1&0 \\0&0&0&1\\ 1&0&0&0\\0&1&0&0  \end{array}\right],\\
&-i \to  [e_4]=\left[\begin{array} {*{20}c}  0&0&0&1\\ 1&0&0&0\\0&1&0&0  \\ 0&0&1&0 \end{array}\right].
\end{split}
\end{equation}
The choise of basis is ambiguous. Any four  $[4 \times 4]$  [0,1]-matrices, which satisfy the rules of cyclic recurrence
\begin{equation}
\label{eq64}
i^1 = i, i^2 = -1, i^3= -i, i^4 = 1
\end{equation}
 can be basis of complex numbers.
 In particular, the following 
$[4 \times 4]$  [0,1]-matrices
\begin{equation}
\label{eq65}
\begin{split}
\raisetag{40pt}
&[e'_1]=\left[\begin{array} {*{20}c}1&0&0&0 \\  0&1&0&0  \\ 0&0&1&0 \\  0&0&0&1\end{array}\right], [e'_2]=\left[\begin{array} {*{20}c}  0&0&1&0  \\ 0&0&0&1 \\0&1&0&0\\ 1&0&0&0 \end{array}\right],\\
&[e'_3]=\left[\begin{array} {*{20}c} 0&1&0&0 \\ 1&0&0&0 \\0&0&0&1 \\ 0&0&1&0 \end{array}\right],
[e'_4]=\left[\begin{array} {*{20}c} 0&0&0&1 \\ 0&0&1&0 \\ 1&0&0&0 \\ 0&1&0&0
\end{array}\right]
\end{split}
\end{equation}
can also be basis of complex numbers. Naturally, the set of [0,1]-matrices, given by (\ref{eq65}) is isomorphous to the set, which is given by (\ref{eq63}).
It is evident, that the system of complex numbers can be constructed by infinite number of the ways, at that cyclic basis can consist of $m$ units, $ m \in N$, starting from three. It is remarkable, that the conformity between complex numbers and matrices is realized by biective mappings. It means, that there is also existing the inverse mapping, by means of which to any   squarte matrix, belonging to the linear space with a basis given by (\ref{eq63}), or (\ref{eq65}), or any other, satisfying the rules of cyclic recurrence like to (\ref{eq64}) can be set up in conformity the complex number. In particular, to any Hermitian  matrix $H$ can be set up in conformity the following complex number
\begin{equation}
\label{eq66}
  \zeta: H \to  S + iA = \left[\begin{array} {*{20}c} S&-A  \\ A&S  \end{array}\right], 
\end{equation}
where $S$  and $A$  are symmetric and antisymmetric parts of Hermitian  matrix.
In the works \cite{Dovlatova_Yerchuck}, \cite{Yerchuck_Dovlatova} was used the apparatus of hypercomplex n-numbers  by
calculation of the optical properties of 1D carbon zigzag shaped nanotubes, taking
into account quantum nature of EM-field. Hypercomplex n-numbers were defined to
be the elements of commutative ring $Z_n$, being to be the direct sum of $n$ fields $C$ of complex numbers, $n\in N$, that is

\begin{equation}
\label{Eq5}
Z_n = C \oplus {C} \oplus{...}\oplus {C}.
\end{equation}
 It means, that any hypercomplex $n$-number $z \in Z_n$ is $n$-dimensional quantity with the components $k_\alpha \in C, \alpha = \overline{0, n-1}$, It can be represented in row matrix form $z \in Z_n$   
\begin{equation}
\label{Eq6}
z = [k_\alpha] = [k_0, k_1, ..., k_{n-1}].
\end{equation}
On the other hand,hypercomplex $n$-number $z \in Z_n$  can also be represented in the form
\begin{equation}
\label{Eq7}
z = \sum_{\alpha = 0}^{n-1}k_\alpha\pi_\alpha,
\end{equation}
where $\pi_\alpha$ are basis elements of $Z_n$. They are the following
\begin{equation}
\label{Eq8}
\begin{split}
&\pi_0 = [1,0, ...,0,0],  \pi_1 =[0,1, ...,0,0],\\
&..., \pi_{n-1} = [0,0, ...,0,1].
\end{split}
\end{equation}
Basis elements $\pi_\alpha$ possess by projection properties
\begin{equation}
\label{Eq9}
\pi_\alpha\pi_\alpha = \pi_\alpha\delta_{\alpha\beta}, \sum_{\alpha = 0}^{n-1}\pi_\alpha = 1, z \pi_\alpha = k_\alpha\pi_\alpha
\end{equation}
Consequently, the set of $k_\alpha \in C, \alpha = \overline{0, n-1}$ is the set of eigenvalues of hypercomplex $n$-number $z \in Z_n$, the set  of $\{\pi_\alpha\}$, $\alpha = \overline{0, n-1}$ is eigenbasis of $Z_n$-algebra.

The simplest hypercomplex number is quaternion.
Any quaternion number $x$ can be determined according to relation
\begin{equation} 
\label{eq1Acacdegh}
\begin{split}
\raisetag{40pt}
&x = (a_1 + ia_2)e + (a_3 + ia_4)j,
\end{split}
\end{equation}
where $\{a_m\}\in R, m = \overline{1,4}$ and
 $e, i, j, k$ produce basis, elements of which are satisfying the conditions  
\begin{equation} 
\label{eq15Acacdegh}
\begin{split}
\raisetag{40pt}\\
&(ij) = k, (ji) = -k, (ki) = j, (ik)= -j,\\
&(ei)= (ie) = i, (ej)= (je) = j, (ek) = (ke) = k.
\end{split}
\end{equation}
Quaternion set \{Q\} can be considered to be the generalization of the set $Z_n$ for n = 2 in accordance with the relation
\begin{equation}
\label{Eq5a}
Z^Q_2 = C e \oplus {C} j,
\end{equation}
that is, it represents itself noncommutative ring. Geometrically    set \{Q\}   can be interpreted to be consisting of two mutually perpendicular  planes in the space $R^Q_4$, determined by orthogonal one real axis and three  imaginary axes, at that i, j , k are imaginary unities  along coordinate axes in three dimensional imaginary $Z^i_3$ space, being to be the subspace of $R^Q_4$. The result of their multiplication is equivalent to $\pi/2$ rotation in the corresponding  pure imaginary plane. It seems to be evident, that the space $R^Q_4$ is isomorphous to Minkowski space $R_4$.
Taking nto account given interpretation, to any quaternion number $x = (a_1 + ia_2)e + (a_3 + ia_4)j$ can be set up in conformity the $[2\times 2]$-matrix according to biective
mapping
\begin{equation}
\label{eq1ABabcd}
\begin{split}
\raisetag{40pt}
&q : (a_1 + ia_2)e + (a_3 + ia_4)j \to\\
&\left[\begin{array} {*{20}c} a_1&-a_2 \\ a_2&a_1 \end{array}\right]E+\left[\begin{array} {*{20}c} a_3&-a_4 \\ a_4&a_3 \end{array}\right]J,
\end{split}
\end{equation}
where E is $[2 \times 2]$ unity matrix, J is the following  $[2 \times 2]$ matrix
\begin{equation}
\label{eqA61}
J = \left[\begin{array} {*{20}c} 0&1  \\ -1&0 \end{array}\right]. 
\end{equation}
 The matrices
\begin{equation}
\label{eqAB61}\begin{split}
\raisetag{40pt}
&E = \left[\begin{array} {*{20}c} 1&0  \\ 0&1 \end{array}\right], 
I = \left[\begin{array} {*{20}c} 0&-i  \\ i&0 \end{array}\right],\\
&J = \left[\begin{array} {*{20}c} 0&1  \\ -1&0 \end{array}\right],
K = \left[\begin{array} {*{20}c} i&0  \\ 0&i \end{array}\right]
\end{split}
\end{equation}
produce basis for  the set $\{x\}$  of quaternion numbers $\{x\} = \{(a_1 + ia_2)e + (a_3 + ia_4)j\}$   in the linear space  of $[2 \times 2]$-matrices, defined over the field of real numbers. It is evident, that the matrices (\ref{eqAB61}) produce anticommutative multiplicative group. It is also clear,  that the system of quaternion numbers like to complex numbers can be constructed by infinite number of the ways.

It is seen, taking into account the first main postulate of  quantum mechanics, 
that for  quantum mechanics and  quantum theory at all, the relation (\ref{eq66})
is especially significant. It allowed in the work  \cite{A_Dovlatova_D_Yerchuck} to formulate the following statement

1.\textit{Any quantumphysical Hermitian operator in Hilbert space, which is set up in conformity to corresponding classical physical quantity, defines two real sets of observables, being to be  connected between themselves and belonging to the field of real numbers (amplitude and phase  sets or the sets of average values and their dispersion), that is the set of complex quantities is determined by Hermitian operator in general case, which allows correctly describe the real noninstantaneous processes}. 

Practical consequence of the statement 1 for quantumphysical tasks is the
following. In particular, by the solution of main quantummechanical equation -
Schroedinger equation,
that is by finding of the eigenvalues ${E_n}$, $n \in N$, of Hamilton operator,
${E_n}$
have to be represented in complex form.
In the application to atom physics ReEn
represents itself the energy of nth atomic
level [more precisely its mathematical
expectation], $ImE_n$ is its dispersion, that is
the width of given level. Usually used solution by $E_n \in R$ leads to the
lost of the half of the possible information.
The conclusion is similar for other
quantumphysical tasks. It seems to be
also remarkable, that the representation of eigenvalues of Hamilton operator in
complex form is used in a practice of
theoretical calculations by a number of authors to be self-evident.

\section{Connection between the symmetry of  differential equations of field theory and main mechanics differential equations  with the mathematical theory of numbers}

Additional gauge invariance of complex relativistic fields was studied in \cite{Yearchuck_Alexandrov_Dovlatova} and in \cite{A_Dovlatova_D_Yerchuck}. It has been found, that conserving quantity, corresponding to invariance of generalized relativistic equations under the operations of additional gauge symmetry group - multiplicative group $\mathfrak R$ of all real numbers (without zero) - is purely imaginary charge. So, it was shown, that complex fields are characterized by complex charges. Let us to remember, that generalized relativistic equations, that is Lagrange equations for any complex relativistic field can be represented in the form of one matrix relativistic differential equation of the first order in partial derivatives. Aalogous equation is taking place for the field  with Hermitian conjugated (complex conjugated in the case of scalar fields) functions $u^{+}(x)$ = $\left\{ {\,u_{i}^{+}(x) \,} \right\}$,$i = \overline{1,n}$. So, the equation for the set $u(x)$ of field functions has the following form
\begin{equation}
\label{eq4Aa}
(\alpha_{\mu} \partial_{\mu} + \kappa \alpha_{0}) u(x) = 0.
\end{equation}
 The equation for the field with Hermitian conjugated (complex conjugated in the case of scalar fields) functions is analogous
\begin{equation}
\label{eq4Ab}
\partial_{\mu} u^{+}(x) \alpha_{\mu} + \kappa u^{+}(x) \alpha_{0} = 0.
\end{equation}

Matrices $\alpha_{\mu}, \alpha_{0}$ in equations (\ref{eq4Aa}, \ref{eq4Ab})  are matrices with constant numerical elements. They have an order, which coincides with the dimension of corresponding space of Lorentz group representation. In particular, they are $[n\times n]$- matrices, if $\left\{ {\,u_{i}(x) \,} \right\}$, $i = \overline{1,n}$ are scalar functions.  Total gauge transformation, established in \cite{Yearchuck_Alexandrov_Dovlatova} and in \cite{A_Dovlatova_D_Yerchuck} is the following
\begin{equation}
\label{eq10w}
u'(x) = \beta exp(i \alpha) u(x),
\end{equation}
where $\alpha,\beta\in R$,  $u(x)$ is the set  of field functions. It is differed from known gauge transformation by presence of factor $\beta\in R$, which is substantial, since it recognizes clearly the additional symmetry of complex relativistic fields.  The

The following statements were proved

A.\textit{Conserving quantity - complex charge, which is invariant under total gauge transformations, corresponds to any complex relativistic field (scalar, vector, spinor).}

B.\textit{Conserving quantity - purely imaginary charge, which is invariant under total gauge transformations, corresponds to any real relativistic field (scalar, vector, spinor).}

 The statement B is especially interesting. It means in fact,  that in complete set of characteristic field functions of any relativistic field (complex or real) enters at least one complex (or pure imaginary) function.

It is understandable, that transformation of field functions by relation  $(\ref{eq10w})$ is equivalent to multiplication of field functions  by arbitrary complex number. The  attention was drawn in \cite{A_Dovlatova_D_Yerchuck}, that the relation  $(\ref{eq10w})$ gives  automorphism of the space of  field functions. It is known, that  automorphism of any linear  space leads to a number useful properties of the objects, which belong to given space. In particular, the consequence of automorphic transformation of the space of EM-field functions by relationship (\ref{eq10w}) is the following. The conservation law for charge will  always take place. At the same time the energy conservation law and the conservation of Poynting vector will be fullfilled, if given transformation is applied to EM-field potentials \cite{Yearchuck_Alexandrov_Dovlatova}, \cite{A_Dovlatova_D_Yerchuck}.    However the energy conservation law and the conservation of Poynting vector, that is mathematical construction, to which enter $\vec{E}$-, $\vec{H}$-vector functions, will not  be fullfilled by transformation (\ref{eq10w}) at arbitrary $\beta$. 
 The charge remains to be Lorentz invariant quantity, at the same time both the field characteristcs, the energy and impulse (determined by Poynting vector) are not Lorentz invariant quantities \cite{A_Dovlatova_D_Yerchuck}. It is remarkable, that the conclusion on charge  Lorentz invariance was formulated in \cite{Tomilchick} to be self-evident. It was concluded in \cite{A_Dovlatova_D_Yerchuck}, that the charge conservation law for EM-field can be considered  to be more fundamental, since it is fullfilled  even through the energy and impulse conservation laws do not take place.

Both the statements give the key for correct generalization of relativistic  field equations, in particular,  electrodynamics equations.

 Given results can be
generalized, if to take into consideration,
that differential equations $(\ref{eq4Aa})$  and $(\ref{eq4Ab})$ are invariant
under the same transformation $(\ref{eq10w})$. Therefore, it is
seen, that the correspondence between the symmetry of differential equations and
 the mathematical nature [in the concept of the number theory] of the
quantities, incoming in given equations seems to be taking place. Given suggestion can be proved and generalized. 

So, we come to the statement:

2.\textit{Differential equations, which are invariant under transformations of groups,
which are symmetry groups of mathematical numbers (considered in the
frame of the number theory) determine the mathematical nature of the quantities,
incoming in given equations}.

In the case of invariance of  differential equations under transformation $(\ref{eq10w})$, that is in the case of invariance under transformations of multiplicative group of complex numbers, the proof seems to be evident. Really,  the multiplication, for instance, of full set of
field function on complex numbers means,
that the functions themselves have to be
complex.

It seems to be essential that differential
equations for dynamics of nonrelativistic classical mechanics
 are invariant under
transformation $(\ref{eq10w})$ too. Really, the dynamics of classical mechanics systems is described by Lagrange equations or by equivalent  canonical Hamilton equations. It is known, that, for instance, Hamilton equations are invariant under contact transformations of variables, that is under the transformation of linear elements - positions and directions, but not points. The transformation $(\ref{eq10w})$ is referred to given class. It means, that by
quantization all physical
quantities, which determine  the dynamics of classical mechanics systems  have to be represented by quantum-mechanical description
 by one of the variants of the
representation of complex number, in
particular, taking into account (\ref{eq66}), by Hermitian matrices.
Therefore, we have the proof of the
statement, being to be the proof of the
main postulate of quantum mechanics.

3.\textit{To any mechanical quantity can be set
up in the correspondence the Hermitian
matrix by quantization.}

It is in fact the consequence of the
statement 2.
The choose of construction of mathemaical apparatus of quantum mechanics on the base of Hermitian matrices is convenient,
however, it is the only one variant from the
infinity of variants of the representations
of quantum mechanical quantities by complex numbers.

We have to remark, that the
description of the processes in classical
mechanics by means of complex number
is also correct. But in very many practical cases, for instance, for mechanical tasks, described by Newton equation, force and impulse can be characterized by the same
phase factor (that is, it can be not taking
into consideration).

Futher will be considered the case of invariance of  differential equations under transformation of anticommutative group of quaternion numbers.
It is the case of electrodynamics.

4.\textit{To any electrodynamics quantity can
be set up in the correspondence the
Quaternion (that is twice-Hermitian)
matrix by quantization of EM-field}.

The symmetry studies of Maxwell equations gave new insight on  the nature of electromagnetic  field. It  has in general case quaternion  structure,
consisting of four independent field constituents, which differ each other by the parities under space inversion
and time reversal. 
 The presence along with vector quaternion characteristics  the independent scalar quaternion characterics of EM-field allows to  describe EM-field instead of unobservable vector and scalar potentials by observable electric field 4-vector-function with the components $E_\alpha(\vec{r},t) = \{E_x(\vec{r},t), E_y(\vec{r},t), E_z(\vec{r},t), i \frac{c \rho_e(\vec{r},t)}{\lambda}\}$ and (or in the case of free EM-field) by means of magnetic field 4-vector-function $H_\mu(\vec{r},t) = \{H_x(\vec{r},t), H_y(\vec{r},t), H_z(\vec{r},t), i \frac{c \rho_m(\vec{r},t)}{\lambda}\}$, where  $i c \rho_e(\vec{r},t)$, $ic\rho_m(\vec{r},t)$ are the $j_4(\vec{r},t)$-component of 4-current density, corresponding to contribution of electric and magnetic component  of charge densities correspondingly, $\lambda$ is  conductivity, which for the case of EM-field propagation in vacuum is  $\lambda_v$ = $\frac{1}{120\pi}$ $(Ohm)^{-1}$. 
Then electric $\vec{j}_e(\vec{r},t)$
 (and magnetic $\vec{j}_g(\vec{r},t)$ in general case) current densities in
the right side of Maxwell equations can be represented
by well known relations $\vec{j}_e(\vec{r},t) = \rho_e(\vec{r},t) \vec{v}_e(\vec{r},t)$, where $\vec{v}_e(\vec{r},t)$is charge velocity in joint system \{EM-field + matter\} and analogous relation (in general
case) for magnetic $\vec{j}_g(\vec{r},t)$ current density. Now the invariance of Maxwell equations under transformations $(\ref{eq10w})$ becomes to be evident. Therefore, take into account the
statement 4, we obtain the independent proof that that
all EM-feld quantities have to be considered minimum
being to be complex quantities (for correct description
of electromagnetic phenomena). Given picture is practically always used both in the theory and in practical
applications. However, even given representation in some
cases is insufficient, it is concerns for instancte the dynamics of optical transitions. At the same time, it is easily to see, that Maxwell equations are invariant under transformations of quaternion non-abelian multiplicative group. In its turn, it leads to conclusion, that really to any electrodynamics quantity can
be set up in the correspondence the
Quaternion (that is twice-Hermitian)
matrix by the quantization of EM-field. It is in fact the consequence of the presence along with symmetry of Maxwell equations under transformations, given by  $(\ref{eq10w})$, the Rainich dual symmetry and additional dual symmetry, established in \cite{A_Dovlatova_D_Yerchuck}.
Let us remark for
comparison, that the equations of the dynamics of mechanical systems are not invariant under transformations
of quaternion multiplicative group. It is the consequence
of non-abelian character of given group. Given conclusion seems to be significant. It means in particular that quantum mechanics in distinction from  quantum electtodynamics cannot be formulated within the frames of quaternion theory. At the same time there are at  present a number of publications, devoted just re-formulation of quantum mechanics on quaternion basis, for instance, \cite{Arbab_I.}, \cite{Arbab_I._Arbab}, \cite{Arbab}, \cite{Arba}. Moreover they are very popular. Let us give the concrete example of incorrectness of given re-formulation. Author of \cite{Arbab_I.}, \cite{Arbab_I._Arbab}, \cite{Arbab}, \cite{Arba} has claimed on the derivation of "universal quantum wave equation" that yields Dirac, Klein-Gordon, Schrodinger 
and quantum heat equations.  Hovever, the author itself indicates that the additional transformations are needed to agree so colled "universal quantum equation" with  Dirac equation. They are the following:  replacing the particle mass $m_0$ by $im_0$, or 
 changing space and time coordinates by $it$ and $i\vec{r}$, respectively. Physical nature of given replacement is not explained by author and it seems to be evident, that it cannot be understandable in the principle. It is mathematical consequence of non-abelian character of multiplcative group of quaternions. The resemblance of the forms of  "universal quantum wave equation"  and Dirac, or Klein-Gordon equations derived from given equation can be explained by minimal deviation of non-abelian character of multiplcative group of quaternions from the abelian groups, it is the only anticommutative. 

An attempt of a new formulation of quantum mechanics based on differential commutator brackets is undertaken in \cite{Faisal}. Differential commutator brackets were introduced in \cite{Arbab_Faisal}. The authors of \cite{Faisal}, \cite{Arbab_Faisal} have used the following relations

\begin{equation}
\label{eq11}
[\frac{\partial}{\partial{t}},\nabla] = 0,
[\frac{\partial}{\partial{t}},\nabla]\cdotp = 0,
[\frac{\partial}{\partial{t}},\nabla] \times = 0,
\end{equation} 
which seems to be correct in application to scalar or vector wave functions. However they have used in subsequent consideration the quaternion wave function, which is unapplicable immediately for the description of quantum mechanics systems. 

So, from given consideration emerges the task: is it possible to built quantum mechanics on quaternion basis, or not. We know, taking into account the conclusions in the Section 2, that there is infinite number of the way to build the quantum theory from the corresponding classical theory, which corresponds to infinite number of the way to build the complex numbers in matrix form. On the other hand, there is the connection between quaternions and complex numbers above considered. It becomes then clear, that new basis for quaternion set has to be choosed, which will be simultaneously the basis of complex numbers. Given task is correct. Really, let us designate the quaternion set by $\{\mathcal{Q}\}$

\begin{equation} 
\label{eq12}
\begin{split}
\raisetag{40pt}
&\{\mathcal{Q}\} = \{Q = a_0 e + a_1 i + a_2 j + a_3 k | a_0, a_1, a_2, a_3 \in R\}.
\end{split}
\end{equation}
$\{\mathcal{Q}\}$ is non-abelian division ring and form a 4-dimensional normed division algebra over
the real numbers $R$. Algebras of the real numbers  $\mathcal{R}$ and complex numbers   $\mathcal{C}$ are subalgebras of  $\{\mathcal{Q}\}$ algebra. Let us designate  the quaternionconjugate
of $Q = a_0 e + a_1 i + a_2 j + a_3 k$  by $Q^*$. Furthermore, define $ReQ:= a_0$ and
 $\vec{Q} = a_1 i + a_2 j + a_3 k$,  that is, usually used biquaternion form will be considered.
 To obtain the quaternion  division ring  with a 4-dimensional normed division algebra over
the real numbers $R$ and with basis, which is simultaneously the basis of complex numbers, it is sufficient, for example, for the quaternion basis \{e, i, j, k \}  to set in the correspondence the matrices, determined by the following bijective mappings 

\begin{equation}
\label{eq13}
\begin{split}
\raisetag{40pt}
& e: \rightarrow E = \left[\begin{array} {*{20}c} 1&0  \\ 0&1 \end{array}\right], 
i: \rightarrow I' =  i \sigma_z = \left[\begin{array} {*{20}c} i&0  \\ 0&-i \end{array}\right],\\
&j: \rightarrow J' = i \sigma_y = \left[\begin{array} {*{20}c} 0&1  \\ -1&0 \end{array}\right],
k: \rightarrow i \sigma_x = K' = \left[\begin{array} {*{20}c} 0&i  \\ i&0 \end{array}\right], 
\end{split}
\end{equation}
where $\sigma_x$, $\sigma_y$, $\sigma_z$  are Pauli matrices. It is easily to see, that the matrices $E$ and $-iJ'$ produce basis of complex numbers in  the quaternion set $\{\mathcal{\hat{Q}}\}$. There is  the biective mapping between quaternion set $\{\mathcal{Q}\}$ and the new quaternion set $\{\mathcal{\hat{Q}}\}$ with basis, determined by (\ref{eq13}).
\begin{equation}
\label{eq14}
\begin{split}
\raisetag{40pt}
&\{Q\}: \rightarrow  \{\mathcal{\hat{Q}}\}: \rightarrow  \{[\mathcal{\hat{Q}}]\} = \{ \left[\begin{array} {*{20}c} a_0 e + i a_1 & a_2  + i a_3   \\  -a_2  + i a_3 &a_0 e - i a_1  \end{array}\right]\},
\end{split}
\end{equation}
where $\{[\mathcal{\hat{Q}}]\}$ is the matrix representation of quaternion set $\{\mathcal{\hat{Q}}\}$, which belongs simultaneously to the representation of complex numbers $\mathcal{C}^{2 \times 2}$. The quaternion conjugates $\{\hat{Q}^*\}$ in given representation are the Hermitian conjugates of the  matrices, determined  by (\ref{eq14}), that is $\{\hat{Q}^*\}$ = $\{[\hat{Q}]^+\}$.  
The quaternion norm is the square root of the determinant
\begin{equation}
\label{eq15}
\|\hat{Q}\| = \sqrt{\det[\hat{Q}]}.
\end{equation}
Mapping, given by (\ref{eq14}), realizes the connection between quaternions and operators on Hilbert spaces.

Let us remark, that, if the quaternions are defined by equations (\ref{eq1Acacdegh}) - (\ref{Eq5a}), then it is more convenient to realise the biective mapping, similar to mapping  (\ref{eq14}) by means of the matrices $E, I, J, K$, determined by relations (\ref{eqAB61}). It is the second way  of the correct formulation of quantum  mechanics  on quaternion base.

Therefore, to quaternion numbers are set up in the correspondence the complex numbers, matrix representations of which is choosed by two ways above indicated, that is, two concrete representations are choosed from the infinite sets of representation forms. It, in its turn, means, that the differential equations, which describe the dynamics of mechanical systems, will be invariant under multiplicative group of the complex numbers, obtained ingiven two ways and, consequently, transformed  sets of quaternion numbers can be used to build the new  two versions of quantum mechanics. 

So, we have obtained the strict proof, that direct application of quaternions with usually used basis \{e, i, j, k \}  to build the new version of quantum mechanics is incorrect and we have found the ways for the creation of the correct versions of quantum mechanics on the quaternion base. 

\section{Conclusions}

The symmetry study of main differential equations of mechanics and electrodynamics  has shown, that differential equations, which
are invariant under transformations of groups, which are symmetry groups of mathematical
numbers (considered within the frames of the number theory) determine the mathematical nature of the
quantities, incoming in given equations. 

The main postulate of quantum mechanics, consisting in that, that to any mechanical
quantity can be set up into the correspondence the Hermitian matrix by quantization was proved.

High symmetry of Maxwell equations, consisting in the presence along with symmetry  under transformations, given by  $(\ref{eq10w})$, the Rainich dual symmetry and additional hyperbolic dual symmetry, established in \cite{A_Dovlatova_D_Yerchuck}, allowed to show, that to EM-field functions, incoming in given equations, can be set up into the correspondence the quaternion (twice-Hermitian) matrices by their quantization.

It is concluded, that the equations of the dynamics of mechanical systems are not invariant under transformations
of quaternion multiplicative group and, consecuently, direct application of quaternions with usually used basis \{e, i, j, k \}  to build the new version of quantum mechanics is incorrect. It is the consequence
of non-abelian character of given group. At the same time  we have found the correct ways for the creation of the new versions of quantum mechanics on the quaternion base by means of choice of new bases in quaternion ring, from which can be formed the bases for complex numbers under multiplicative groups  of which the equations of the dynamics of mechanical systems are  invariant.

\end{document}